\documentclass{PoS}

\title{Search for Flavor Changing Neutral Current in Top Production and Decays}

\ShortTitle{}

\author{Tae Jeong Kim on behalf of the CMS collaboration\\
        Hanyang University, Seoul, South Korea\\
        E-mail: \email{taekim@hanyang.ac.kr}}

\abstract{
Searches for flavor changing neutral currents in top production and decay using data collected by
the Compact Muon Solenoid (CMS) experiment at $\sqrt{s}$ = 7 and \mbox{8 TeV} are presented,
corresponding to an integrated luminosity of around 5 fb$^{-1}$ and 20 fb$^{-1}$. 
FCNC searches are conducted to probe $tqZ$, $tq\gamma$, $tqH$, and $tgq$ interactions in various channels.
By the time of the 38$^{th}$ ICHEP conference in 2016, 
the upper limits on \mbox{$\mathcal{B}(t \to u\gamma)$ $<$ 0.013\% }, \mbox{$\mathcal{B}(t \to ug)$ $<$ 0.036\%}, \mbox{$\mathcal{B}(t \to uZ)$ $<$ 0.05\%} and
\mbox{$\mathcal{B}(t \to uH)$ $<$ 0.42\%} at the 95\% confidence level had been obtained by the CMS collaboration.
}

\FullConference{38th International Conference on High Energy Physics\\
		3-10 August 2016\\
		Chicago, USA}

\begin{document}

\section{Introduction}
In the standard model (SM), the Flavor Changing Neutral Current (FCNC) can occur only at the level of quantum loop
correction with the branching ratio of less than 10$^{-13}$. However, a wide variety of models beyond the SM shows a strong
dependence in the measurable FCNC quantities, e.g. the branching ratio of top decaying to a charm quark and a Higgs boson in 2HDM 
is around 10$^{-4}$. Therefore, the study of FCNC is one of the most interesting research topic in top quark physics.
The Compact Muon Solenoid (CMS) experiment~\cite{CMSdet} at the LHC has accumulated data corresponding to an integrated luminosity
of around 5 fb$^{-1}$ at $\sqrt{s}$ = 7 TeV in 2011 and 20 fb$^{-1}$ at $\sqrt{s}$ = 8 TeV in 2012.
Searches for FCNC in top production and decay using
data collected by the CMS experiment at the time of the ICHEP 2016 conference 
are presented. FCNC searches are conducted to probe $tqZ$, $tq\gamma$, $tqH$, and $tgq$ interactions in various channels.

\section{Search for \bf $tqZ$ coupling in top decay}
\begin{figure}[h]
\begin{minipage}{18pc}
\centering
\includegraphics[width=14pc]{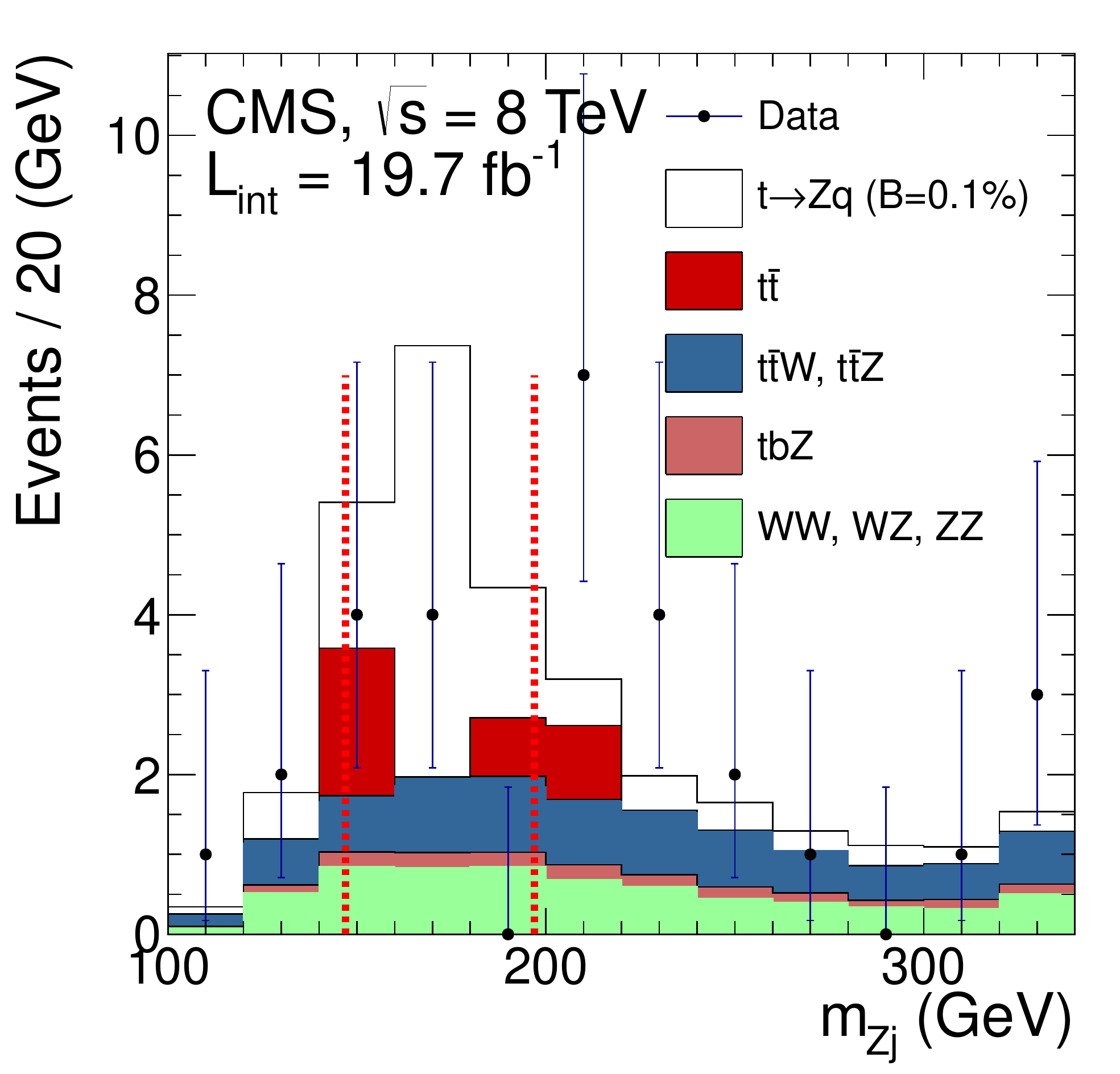}
\end{minipage}\hspace{0pc}%
\begin{minipage}{18pc}
\centering
\includegraphics[width=14pc]{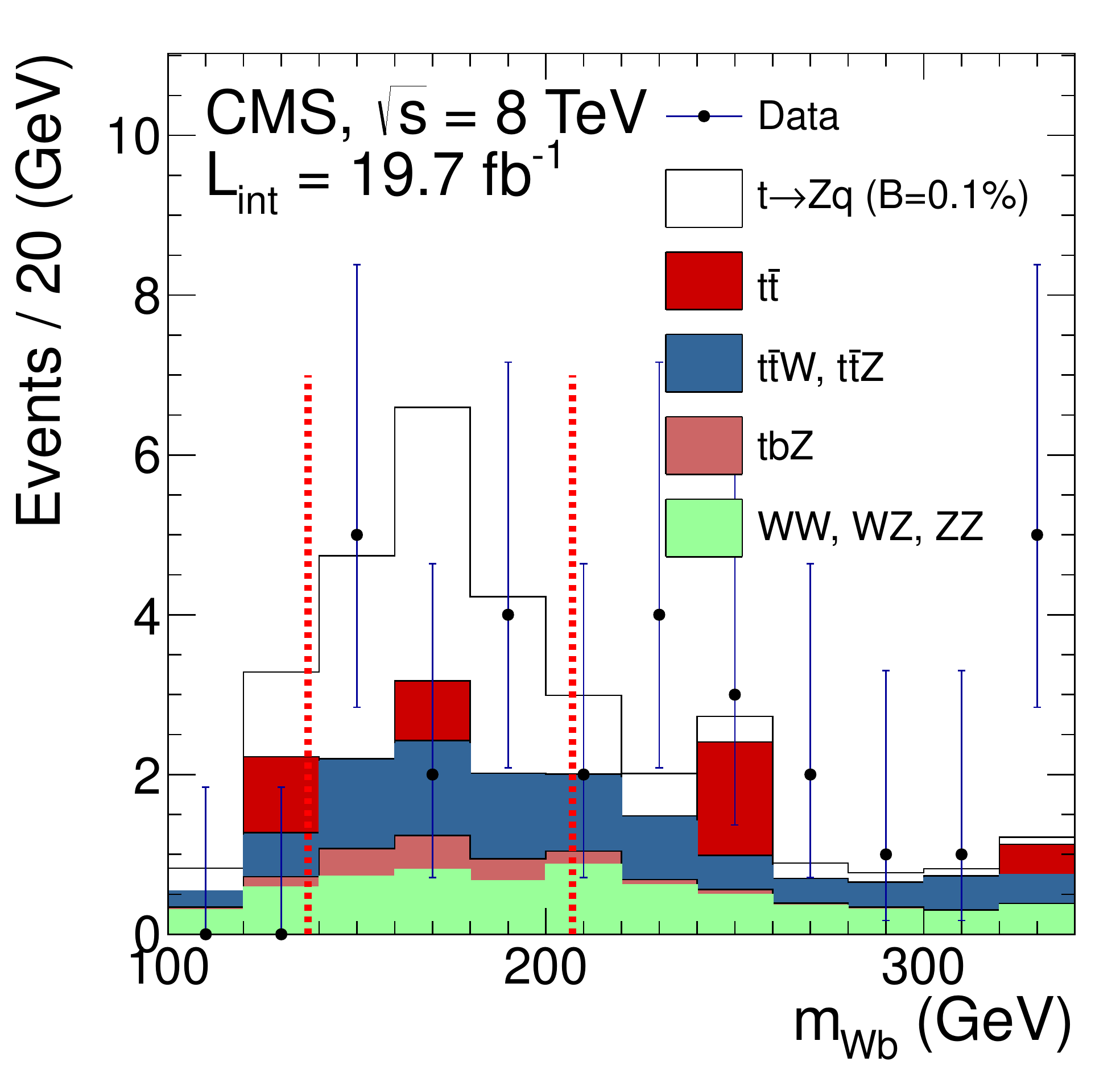}
\end{minipage}\hspace{0pc}%
\caption{\label{fig:fig1}
The distributions of reconstructed top quark mass in the SM decay (left) and FCNC decay (right) are shown. 
}
\end{figure}

Top-quark pair events where one of top decays to a Z boson and a up-type quark have been analyzed searching for
the coupling of $tqZ$ with data corresponding to an integrated luminosity of 4.9 fb$^{-1}$ collected 
at 7 TeV and 19.7 fb$^{-1}$ at 8 TeV~\cite{RefFCNCtopdecay}. 
Events are selected with three leptons ($e, \mu$): two leptons from a Z boson and one lepton from a W boson in top decays.  
The distribution of reconstructed top quark mass in the SM decay (left) and FCNC decay (right) are 
shown in Fig.~\ref{fig:fig1} which show a good agreement between data and the SM background prediction.
An observed limit on the branching ratio of $\mathcal{B}(t \to Zq) < 0.5\%$ is obtained.

\section{Search for \bf $tqH$ coupling in top decay}
\begin{figure}[h]
\begin{minipage}{18pc}
\centering
\includegraphics[width=14pc]{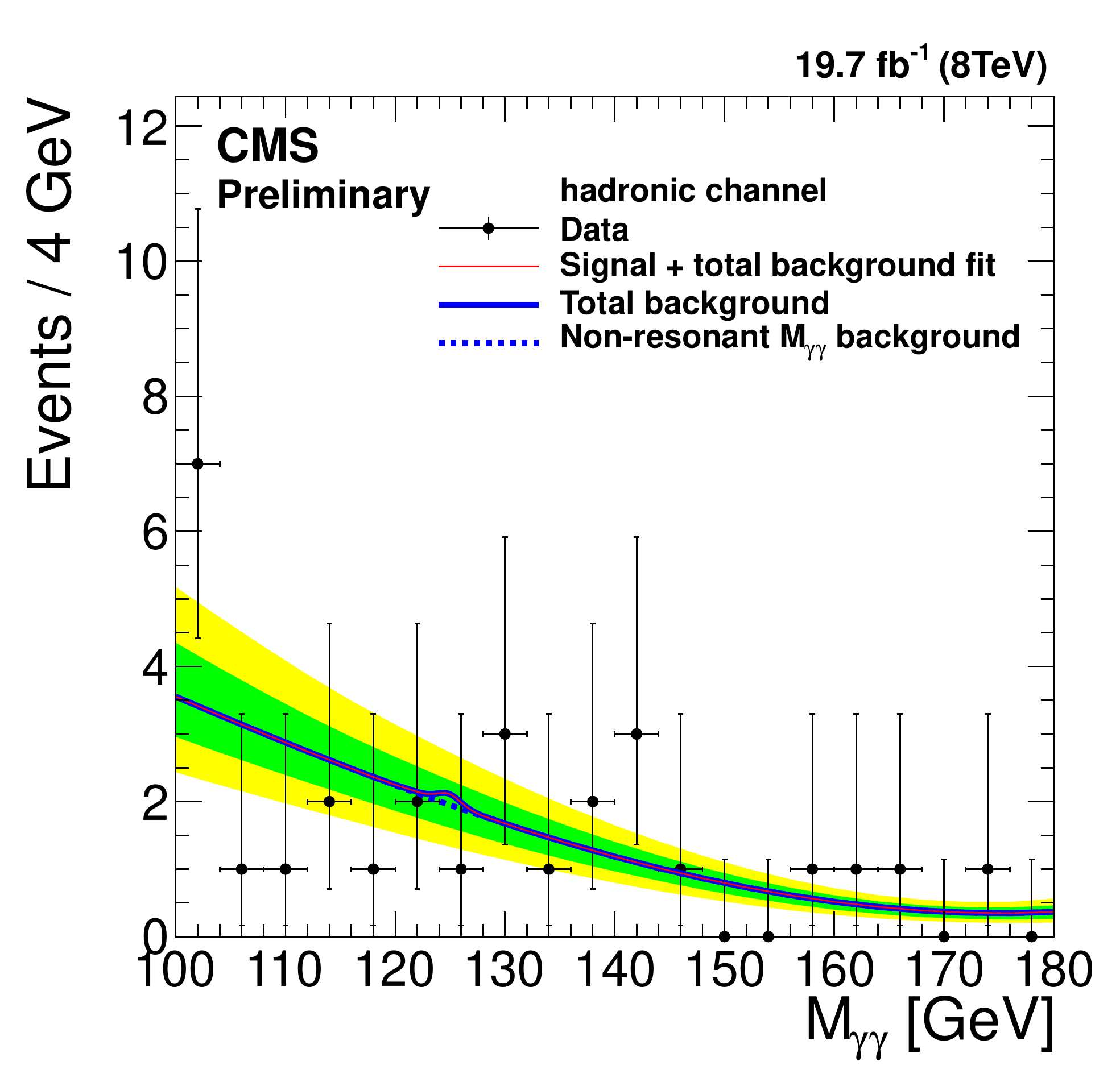}
\end{minipage}\hspace{0pc}%
\begin{minipage}{18pc}
\centering
\includegraphics[width=14pc]{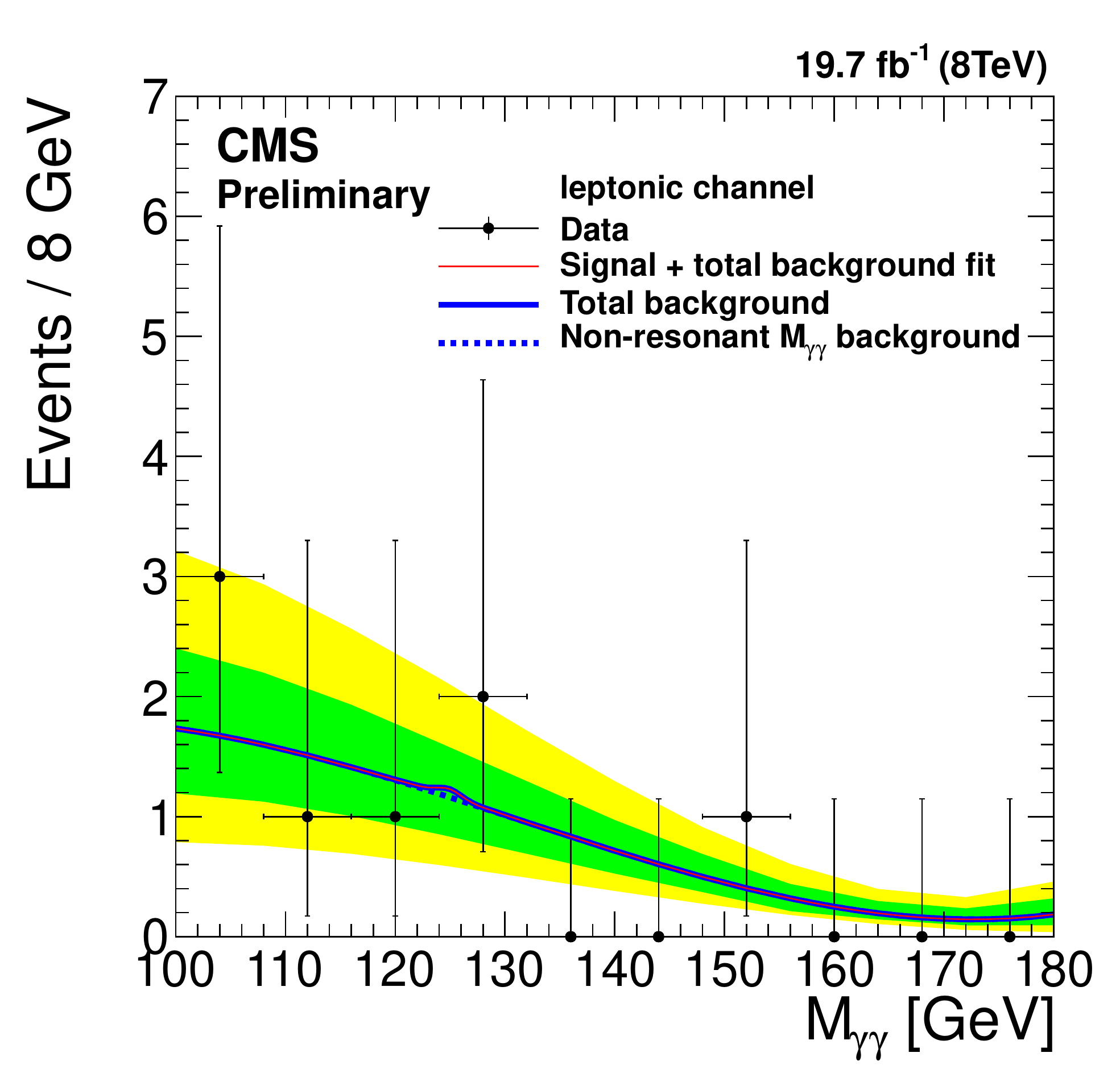}
\end{minipage}
\caption{\label{fig:fig2}
Diphoton invariant mass distributions and fit result for the hadronic channel (left) and the leptonic channel (right).
The blue solid line indicates the total background including the resonant from the Higgs boson and the non-resonant diphoton background. 
}
\end{figure}

The search for the rare flavor-changing decay of the top quark to a Higgs boson and a charm or up quark~\cite{RefTOP-14-019,RefTOP-13-017,RefTOP-14-020} 
have also been performed.
Top-quark pair events in the hadronic and leptonic channel, where the Higgs boson decays two photons are used in this search. 
Figure~\ref{fig:fig2} shows the diphoton invariant mass distributions for hadronic channel (left) and leptonic (right) channel.
The $\gamma\gamma$ resonant from the SM Higgs is taken into account in the background modeling.   
No excess over the data is found in both analysis, being the sensitivity driven by the hadronic channel. 
The result from the final state of diphoton search analysis combining two channels is $\mathcal{B}(t \to cH)$ $<$ 0.47\% and
is 0.42\% for $\mathcal{B}(t \to uH)$. The main uncertainties are from the photon identification and non-resonant background estimation. 

\section{Search for \bf $tq\gamma$ coupling in single top production}
\begin{center}
\begin{figure}[h]
\begin{minipage}{18pc}
\centering
\includegraphics[width=10pc]{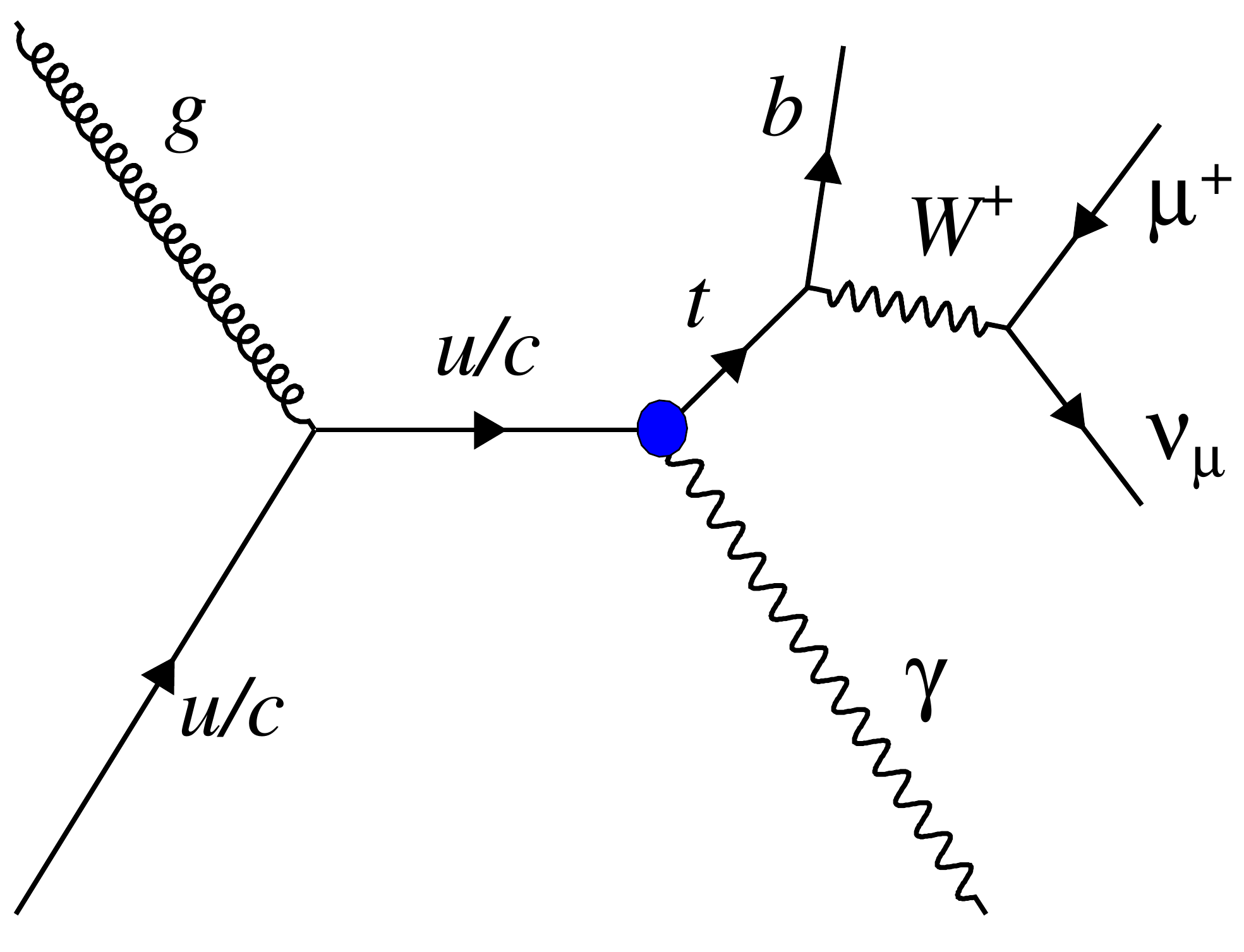}
\end{minipage}\hspace{0pc}%
\begin{minipage}{18pc}
\centering
\includegraphics[width=16pc]{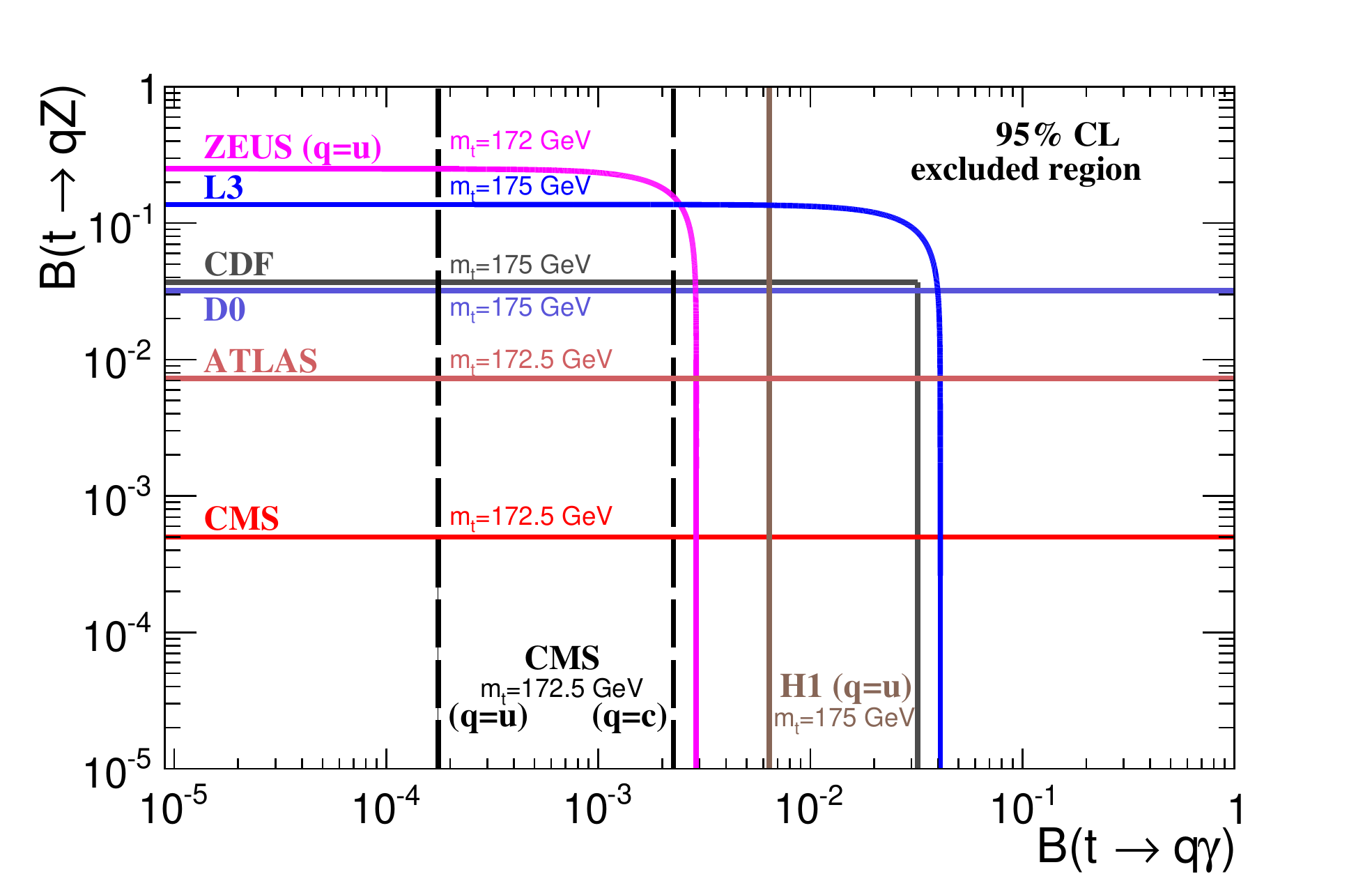}
\end{minipage}
\caption{\label{fig:fig3}
(Left) Lowest-order Feynman diagram for single top quark production in association with a photon via a FCNC. 
(Right) The measured 95\% CL upper limits on \mbox{$\mathcal{B}(t \to qZ)$} versus \mbox{$\mathcal{B}(t \to q\gamma)$}. 
The two vertical dashed lines show the results of this analysis for the $t \to q\gamma$ coupling. 
}
\end{figure}
\end{center}

The anomalous $tq\gamma$ coupling has been searched with events in association with a photon using the full data collected at 8 TeV in 2012, 
corresponding to 19.1 fb$^{-1}$~\cite{RefTOP-14-003}.
The corresponding Feynman diagram is shown in Fig~\ref{fig:fig3} (left).
In order to reduce the QCD multijet background, the top quark in the lepton decay mode is considered.
Therefore, one isolated muon, one isolated photon and one $b$-tagged jet are required to select the final signature.
A Boosted Decision Tree (BDT) is used to separate the signal signature from the background contributions.
There is no excess observed.
Photon energy scale and the estimate of $W\gamma$ + jets process are the main uncertainty sources in this analysis.
The upper limits on the branching ratios of $tu\gamma$ and $tc\gamma$ is
shown in Fig~\ref{fig:fig3} (right), being the case of coupling with c-quark and u-quark indicated by red vertical line.
The observed limits on the branching ratios of $\mathcal{B}(t \to u\gamma)$ $<$ 0.013 \% and $\mathcal{B}(t \to c\gamma)$ $<$ 0.17 \% are obtained.
This result is the first measurement of the limit on these couplings at the LHC and provide the most stringent bounds on the anomalous FCNC $tq\gamma$ coupling to date.

\section{Search for \bf $tqg$ coupling in single top production}\label{sec:tqg}
The anomalous couplings of $tcg$ and $tug$ were searched in the $t$-channel single top-quark production using data collected at 7 TeV
corresponding to an integrated luminosity of 5 fb$^{-1}$~\cite{RefTOP-14-007}.
The final signature of the anomalous FCNC coupling is the same as the SM single top quark production processes
with one isolated muon, 2 or 3 jets and one $b$-tagged jet.
However, the signal has different kinematic distributions and is
separated from the backgrounds using
a Bayesian Neural Network (BNN) based on the distinctive kinematical properties.
Data and the SM predictions agree well within the uncertainties.
In this analysis, one of the main sources of uncertainty is the estimate of the W+jets contribution to the signal region.
Taking into account the systematic uncertainties, the limit on the couplings of
$tcg$ and $tug$
are calculated at the 95\% confidence level.
The observed upper limits of $\mathcal{B}(t \to ug) < 0.036 \% $ and $\mathcal{B}(t \to cg) < 0.34 \% $ are obtained.

\section{Conclusions}

The CMS collaboration has performed several FCNC searches
in both top-quark pair and a single top-quark production and decay, with data collected at 7 and 8 TeV.
No sign of FCNC contribution in the top sector has been found so far. Limits on the branching ratios 
\mbox{$\mathcal{B}(t \to u\gamma)$ $<$ 0.013\% }, 
\mbox{$\mathcal{B}(t \to ug)$ $<$ 0.036\%}, 
\mbox{$\mathcal{B}(t \to Zq)$ $<$ 0.05\%} and
\mbox{$\mathcal{B}(t \to uH)$ $<$ 0.42\%}
at 95\% confidence level are obtained.
More results from Run 1 are expected to come soon.
Exciting time are ahead of us with data to be collected at 13 TeV,
which will allow the LHC experiments to probe further the contributions from rare FCNC processes in the top sector. 
After the 38$^{th}$ ICHEP conference, more results~\cite{cms:topresults} on the FCNC became available.


\begin{thebibliography}{99}
\bibitem{CMSdet}
CMS Collaboration, ``The CMS experiment at the CERN LHC", JINST 3:S08004,2008.
\bibitem{RefFCNCtopdecay}
CMS Collaboration, ``Search for Flavor-Changing Neutral Currents in Top-Quark Decays $t \to Zq$ in $pp$ Collisions at
$\sqrt{s}$ = 8 TeV",
Phys. Rev. Lett., {\bf 112} (2014) 171802.
\bibitem{RefTOP-14-019}
CMS Collaboration,
"Search for top quark decays $t \to qH$ with $H \to \gamma\gamma$ in pp collisions at $\sqrt{s}$ = 8 TeV $pp$",
\textbf{PAS TOP-14-019}.
\bibitem{RefTOP-13-017}
CMS Collaboration,
"Search for top quark decays via Higgs-boson-mediated flavor changing",
\textbf{PAS TOP-13-017, arXiv:1610.04857}.
\bibitem{RefTOP-14-020}
CMS Collaboration,
"Search for the flavor-changing neutral current decay $t \to qH$ where the Higgs decays to $b\bar{b}$ pairs at $\sqrt{s}$ = 8 TeV", 
\textbf{PAS TOP-14-020}.
\bibitem{RefTOP-14-003}
CMS Collaboration,
"Search for anomalous single top quark production in association with a photon in pp collisions at $\sqrt{s}$ = 8 TeV",
\textbf{JHEP 04 (2016) 035}.
\bibitem{RefTOP-14-007}
CMS Collaboration,
"Search for anomalous Wtb couplings and top FCNC in t-channel single-top-quark events",
\textbf{PAS TOP-14-007}.
\bibitem{cms:topresults}
https://twiki.cern.ch/twiki/bin/view/CMSPublic/PhysicsResultsTOP.
\end{thebibliography}
\end{document}